\documentclass[12pt]{article}
\usepackage[a4paper, total={7in, 10in}]{geometry}
\usepackage[parfill]{parskip}
\usepackage{physics, tensor, float, subcaption}
\usepackage{graphicx}
\graphicspath{ {Plots/} }
\usepackage{jhep-mod}
\usepackage{bm}
\usepackage{soul}
\usepackage{amssymb,amsmath,amsthm}
\usepackage{mathrsfs}
\usepackage[utf8]{inputenc}
\usepackage{enumerate}
\usepackage{bigints}
\usepackage{xcolor}
\usepackage{appendix}
\usepackage{graphicx}
\usepackage{float}
\usepackage{tikz}
\usepackage{setspace}
\usepackage{cancel}
\usepackage{doi}
\definecolor{purple}{rgb}{1,0,1}

\definecolor{lime}{HTML}{A6CE39}
\newcommand{\orcidicon}{%
	\begin{tikzpicture}
	\draw[lime, fill=lime] (0,0) 
		circle [radius=0.16] 
		node[white] {{\fontfamily{qag}\selectfont \tiny ID}};
	\draw[white, fill=white] (-0.0625,0.095) 
		circle [radius=0.007];
	\end{tikzpicture}
	\hspace{-5mm}
}
\newcommand\orcidJosh{{\href{https://orcid.org/0000-0003-1200-7261}{\orcidicon}}}
\newcommand\orcidThomas{{\href{https://orcid.org/0000-0002-0314-4136}{\orcidicon}}}
\newcommand\orcidAlex{{\href{https://orcid.org/0000-0002-1763-3563}{\orcidicon}}}
\newcommand\orcidMatt{{\href{https://orcid.org/0000-0003-1088-6485}{\orcidicon}}}
\begin{document}


\title{\vspace{-25pt}\huge{
Killing tensor and Carter constant for Painlev\'e--Gullstrand form of Lense--Thirring spacetime
}}


\author{
\Large
Joshua Baines\!\orcidJosh, Thomas Berry\!\orcidThomas, Alex Simpson\!\orcidAlex,\! 
\\{\sf  and} Matt Visser\!\orcidMatt}
\affiliation{School of Mathematics and Statistics, Victoria University of Wellington, 
\\
\null\qquad PO Box 600, Wellington 6140, New Zealand.}
\emailAdd{joshua.baines@sms.vuw.ac.nz}
\emailAdd{thomas.berry@vuw.ac.nz}
\emailAdd{alex.simpson@sms.vuw.ac.nz}
\emailAdd{matt.visser@sms.vuw.ac.nz}

\abstract{
\vspace{1em}

Recently, the authors have formulated and explored a novel Painlev\'{e}--\-Gull\-strand variant of the Lense--Thirring spacetime, which has some particularly elegant features --- including unit-lapse, intrinsically flat spatial 3-slices, and some particularly simple geodesics, the ``rain'' geodesics. 
At linear level in the rotation parameter this spacetime is indistinguishable from the usual slow-rotation expansion of Kerr.
Herein, we shall show that this spacetime possesses a nontrivial  Killing tensor, implying separability of the Hamilton--Jacobi equation.
Furthermore, we shall show that the Klein--Gordon equation is also separable on this spacetime. 
However, while the Killing tensor has a 2-form square root, we shall see that this 2-form square root of the Killing tensor is \emph{not} a Killing--Yano tensor.  Finally, the Killing-tensor-induced  Carter constant is easily extracted, and now, with a fourth constant of motion, the geodesics become (in principle) explicitly integrable. 

\bigskip
\noindent
{\sc Date:} Tuesday 5 October 2021; Monday 18 October 2021; \\
\null\qquad\qquad \LaTeX-ed \today


\bigskip
\noindent{\sc PhySH:} 
Gravitation
}

\maketitle
\def\tr{{\mathrm{tr}}}
\def\diag{{\mathrm{diag}}}
\def\cof{{\mathrm{cof}}}
\def\pdet{{\mathrm{pdet}}}
\def\d{{\mathrm{d}}}
\def\K{{\mathcal{K}}}
\newcommand{\C}{\mathcal{C}}
\parindent0pt
\parskip7pt

\section{Introduction}\label{Intro}

Recently the current authors have introduced and explored a new variant of the Lense--Thirring spacetime~\cite{4-of-us}, specified by the line element
\begin{equation}
\label{E:pglt}
\d s^2 = - \d t^2 +\left\{\d r+\sqrt{\frac{2m}{r}} \; \d t\right\}^2
+ r^2 \left\{\d\theta^2+\sin^2\theta\; \left(\d\phi - {2J\over r^3} \d t\right) ^2\right\} \ .
\end{equation}
The metric components are easily read off as
\begin{equation}
g_{ab} = \left[ \begin{array}{c|ccc}
-1+{2m\over r} + {4 J^2\sin^2\theta\over r^4} &  \sqrt{2m\over r} & 0 & -{2J\sin^2\theta\over r}\\
\hline
\sqrt{2m\over r} & 1 & 0 & 0\\
0& 0 & r^2 & 0\\
-{2J\sin^2\theta\over r} & 0 & 0 & r^2\sin^2\theta\\
\end{array}
\right]_{ab}.
\end{equation}
It is easy to verify that $\det(g_{ab}) = - r^4 \sin^2\theta$, and that the inverse metric is:
\begin{equation}
\label{E:contra_metric}
g^{ab} = \left[ \begin{array}{c|ccc}
-1&  \sqrt{2m\over r} & 0 & -{2J\over r^3}\\
\hline
\sqrt{2m\over r} & 1-{2m\over r} & 0 & \sqrt{2m\over r} \;{2J\over r^3}\\
0& 0 &\; {1\over r^2}\; & 0\\
-{2J\over r^3} &  \sqrt{2m\over r}\; {2J\over r^3} & 0 & {1\over r^2\sin^2\theta} - {4J^2\over r^6}\\
\end{array}
\right]^{ab}.
\end{equation}
This variant of the Lense--Thirring spacetime is rather useful since the metric is recast into Painlev\'{e}--Gullstrand form~\cite{painleve1,painleve2,gullstrand,poisson}. Writing the metric in this form gives it two very useful properties: the first is the property of unit-lapse, characterised by $g^{tt}=-1$, and the second is the possession of a flat spatial $3$-metric,  notably
\begin{equation}
g_{ij} \; \d x^i \,\d x^j \longrightarrow \d r^2 + r^2(\d\theta^2+\sin^2\theta\, \d\phi^2).
\end{equation}
A~flat $3$-metric allows for an almost trivial analysis of the constant-$t$ spatial hypersurfaces, while lapse unity permits straightforward calculation of particular geodesics of the spacetime. Specifically, the ``rain" geodesics become almost trivial to calculate~\cite{Unit-lapse}.
At linear level in the rotation parameter this spacetime is indistinguishable from the usual slow-rotation expansion of Kerr.

We also note the advantages of using this variant of the Lense--Thirring spacetime, as opposed to the exact Kerr solution, in some astro\-physically interesting contexts. Firstly, since there is no analogue of the Birkhoff theorem for axi\-symmetric spacetimes in $(3+1)$ dimensions~\cite{Birkhoff, Jebsen, Deser, Ravndal, Skakala}, the Kerr solution need not, (\emph{and typically will not}), perfectly model rotating horizonless astrophysical sources (such as stars, planets, \emph{etc.}). This is due to the nontrivial mass multipole moments that these objects typically possess. Instead, the Kerr solution will model the gravitational field in the asymptotic regime, where Lense--Thirring serves as a valid approximation to Kerr~\cite{Lense-Thirring, Pfister, Adler-Bazin-Schiffer, MTW, Wald, weinberg, Hobson, D'Inverno, Hartle, Carroll, kerr-intro, kerr-book, Kerr, Kerr-Texas, kerr-newman, kerr-book-2,river}. Secondly, the Lense--Thirring metric is algebraically \emph{much} simpler than the Kerr metric, making most calculations  significantly easier to conduct. 
Furthermore, the Lense--Thirring metric can be recast into Painlev\'{e}--Gullstrand form, while the Kerr metric cannot \cite{Kerr-Darboux,Valiente-Kroon:2004a,Valiente-Kroon:2004b,Jaramillo:2007}.

Given that this variant of the Lense--Thirring metric is amenable to significantly more tractable mathematical analysis, and is a valid approximation for the gravitational fields of rotating stars and planets in the same regime as the Kerr solution is appropriate, there is a compelling argument to use the Painlev\'{e}--Gullstrand form of Lense--Thirring to model various astrophysically interesting cases~\cite{Carballo-Rubio:2018,Visser:2009a,Visser:2009b}. 

Supplementary to this, we will show below that this spacetime possesses a  nontrivial Killing tensor, and we shall also present the 2-form square root of this Killing tensor, an object that acts as a ``would-be'' Killing--Yano tensor. We discuss precisely how this object does and does not satisfy the \emph{desiderata} for being a genuine Killing--Yano tensor. We establish why this candidate spacetime does not possess the full Killing tower (consisting of principal tensor, Killing--Yano tensor, and Killing tensor).
We also check that the Klein--Gordon equation is separable on this variant of Lense--Thirring spacetime.

\enlargethispage{30pt}
Given only three constants of motion: the energy $E$, angular momentum $L$, and particle mass parameter $\epsilon$, the geodesic equations are not integrable. By finding a nontrivial Killing tensor for the spacetime, we generate a fourth constant of the motion, a generalization of the Carter constant $\mathcal{C}$. 

The existence of this additional constant of motion then implies complete separability of the Hamilton--Jacobi equation, which makes the geodesic equations fully integrable, at least in principle.

\vspace{-10pt}
\section{Killing Tensor}
\enlargethispage{20pt}
Nontrivial Killing tensors are incredibly useful mathematical objects that are present in almost all (useful) candidate spacetimes, and can be thought of as generalisations of Killing vectors. A Killing tensor is a completely symmetric tensor of type $(0, l)$ which satisfies the following equation:
\begin{equation}
\nabla_{(b}K_{a_1...a_l)}=0 \ .
\end{equation}
However, unlike Killing vectors, Killing tensors do not naturally arise from explicit symmetries present in the spacetime. Hence finding nontrivial Killing tensors in a spacetime can be difficult in the abstract. However, in two recent papers by Papadopoulos and Kokkotas~\cite{Papadopoulos:2018, Papadopoulos:2020}, which are in turn based on older results by Benenti and Francaviglia~\cite{Benenti:1979},  it has been explicitly shown that if the inverse metric of a spacetime can be written in a particular form, then a nontrivial (contravariant) Killing tensor of rank 2 exists and can be easily calculated.  (Here we make the distinction of requiring a nontrivial Killing tensor since the metric itself is always a trivial Killing tensor). 

To use this method we first coordinate transformed our Lense--Thirring metric variant into Boyer--Lindquist form~\cite{4-of-us}
\begin{equation}
\label{E:BL-form}
(\d s^2)_{BL} = - (1-2m/r)\d t^2 +{\d r^2\over 1-2m/r}
+ r^2 \left\{\d\theta^2+\sin^2\theta\; \left(\d\phi - {2J\over r^3} \d t\right) ^2\right\} \ .
\end{equation}
Here
\begin{equation}
\label{E:bl-g}
(g_{ab})_{BL} = \left[ \begin{array}{c|cc|c}
-1+{2m\over r} + {4 J^2\sin^2\theta\over r^4} &  0 & 0 & -{2J\sin^2\theta\over r}\\
\hline
0 & {1\over1-2m/r} & 0 & 0\\
0& 0 & r^2 & 0\\ 
\hline
-{2J\sin^2\theta\over r} & 0 & 0 & r^2\sin^2\theta\\
\end{array}
\right]_{ab},
\end{equation}
and
\begin{equation}
\label{E:bl-inverse}
(g^{ab})_{BL}  = \left[ \begin{array}{c|cc|c}
-{1\over 1-2m/ r} &  0 & 0 & -{2J\over r^3(1-2m/r)}\\
\hline
0 & 1-{2m\over r} & 0 & 0\\
0& 0 & {1\over r^2} & 0\\
\hline
-{2J\over r^3(1-2m/r)} & 0 & 0 & {1\over r^2\sin^2\theta} - {4J^2\over r^6(1-2m/r)}\\
\end{array}
\right]^{ab}.
\end{equation}

\enlargethispage{30pt}
We then applied the Papadopoulos--Kokkotas algorithm~\cite{Papadopoulos:2018, Papadopoulos:2020}, by first inverting the Boyer--Lindquist form of the metric (\ref{E:bl-g}) to obtain (\ref{E:bl-inverse}), then extracting the contravariant Killing tensor in these coordinates, and finally converting the result back to Painlev\'e--Gullstand coordinates.

\clearpage
After conversion back to Painlev\'e--Gullstand coordinates, where the line element is again~(\ref{E:pglt}), 
the Papadopoulos--Kokkotas algorithm~\cite{Papadopoulos:2018, Papadopoulos:2020}  yields the particularly simple contra\-variant form of the Killing tensor:
\begin{equation}
K^{ab}=
\begin{bmatrix}
0& 0 & 0 & 0\\
0 & 0 & 0 & 0\\
0 & 0 &\; 1 \;& 0\\
0 & 0 & 0 & {1\over\sin^2\theta}
\end{bmatrix}
^{ab} \ .
\end{equation}
The corresponding covariant form of the Killing tensor,  
$K_{ab}= g_{ac} \,K^{cd}\, g_{db}$, is then
\begin{equation}
\label{PGLT_Killing_Tensor}
K_{ab}=
\begin{bmatrix}
\frac{4J^2\sin^2\theta}{r^2} & 0 & 0 & -2Jr\sin^2\theta\\
0 & 0 & 0 & 0\\
0 & 0 & \; r^4\;  & 0\\
-2Jr\sin^2\theta & 0 & 0 & r^4\sin^2\theta
\end{bmatrix}
_{ab} \ .
\end{equation}
One can easily explicitly check that $\nabla_{(c}K_{ab)}= K_{(ab;c)} = 0$, hence equation \eqref{PGLT_Killing_Tensor} does indeed represent a Killing tensor. We can also compactly write:
\begin{equation}
K_{ab} \; \d x^a \; \d x^b = 
r^4\left\{\d\theta^2 + \sin^2\theta \left(\d\phi - {2J\over r^3} \d t\right)^2 \right\}.
\end{equation}
We now adopt an orthonormal basis, using the co-tetrad and tetrad developed in reference~\cite{4-of-us}. For the co-tetrad we take
\begin{eqnarray}
e^{\hat t}{}_a &=& (1;0,0,0); \qquad 
\; \; e^{\hat r}{}_a = \left(\sqrt{2m\over r};1,0,0\right); 
\nonumber\\[2pt]
e^{\hat \theta}{}_a&=& r(0;0,1,0); \qquad 
e^{\hat \phi}{}_a = r\sin\theta\left(-{2J\over r^3};0,0,1\right).
\end{eqnarray}\enlargethispage{20pt}
The corresponding tetrad is then
\begin{eqnarray}
e_{\hat t}{}^a &=& \left(1;-\sqrt{2m\over r},0,{2J\over r^3}\right); \qquad 
\; \; e_{\hat r}{}^a = \left(0;1,0,0\right); 
\nonumber\\[2pt]
e_{\hat \theta}{}^a&=& {1\over r}\;(0;0,1,0); \qquad 
e_{\hat \phi}{}^a = {1\over r\sin\theta}\left(0;0,0,1\right).
\end{eqnarray}
For the tetrad components of the Killing tensor 
we find
\begin{equation}
K_{\hat a\hat b}\longrightarrow r^2
\begin{bmatrix}
0 & 0 & 0 & 0\\
0 & 0 & 0 & 0\\
0 & 0 & 1 & 0\\
0 & 0 & 0 & 1
\end{bmatrix}_{\hat a\hat b}.
\end{equation}
Since this is diagonal, and since we also know from reference~\cite{4-of-us} that the orthonormal form of the Ricci tensor $R_{\hat a\hat b}$ is diagonal, it follows that the Ricci tensor commutes with the Killing tensor: $R^{\hat a}{}_{\hat b} \,K^{\hat b}{}_{\hat c} = K^{\hat a}{}_{\hat b} \,R^{\hat b}{}_{\hat c}$.  Indeed, even in a coordinate basis it follows that $R^a{}_b \,K^b{}_c = K^a{}_b \,R^b{}_c$. Note that the commutator $[R,K]_{ab} = R_{ac} g^{cd} K_{db} - K_{ac} g^{cd} R_{db}$ can be viewed as a $2$-form. It is also potentially useful to note that the trace of the Killing tensor is particularly simple; $K = K^{ab} g_{ab} =  K_{ab} g^{ab} = 2 r^2$.

If we now take the limit $J\rightarrow 0$, then the Lense--Thirring spacetime reduces to the spherically symmetric Schwarzschild spacetime. In this $J\to0$ limit, the nontrivial (covariant) Killing tensor becomes
\begin{equation}
K_{ab}\longrightarrow 
\begin{bmatrix}
0 & 0 & 0 & 0\\
0 & 0 & 0 & 0\\
0 & 0 & \;r^4\; & 0\\
0 & 0 & 0 & r^4\sin^2\theta
\end{bmatrix}
_{ab} \ ,
\end{equation}
so that 
\begin{equation}
K_{ab} \; \d x^a \; \d x^b \longrightarrow
r^4\left\{\d\theta^2 + \sin^2\theta\; \d\phi^2 \right\}.
\end{equation}
\enlargethispage{30pt}
Indeed, it is easily verified that this is the appropriate Killing tensor in any arbitrary spherically symmetric spacetime,  even if it is time dependent. Furthermore, for any arbitrary (possibly time dependent) spherically symmetric spacetime one can always block diagonalize the metric and Ricci tensors in the form
\begin{equation}
g_{ab}\longrightarrow 
\left[
\begin{array}{cc|cc}
\,* & * & 0 & 0\\
\,* & * & 0 & 0\\
\hline
0 & 0 & \;r^2\; & 0\\
0 & 0 & 0 & r^2\sin^2\theta
\end{array}
\right]
_{ab};\qquad\qquad
R_{ab}\longrightarrow 
\left[\begin{array}{cc|cc}
\,* & * & 0 & 0\\
\,* & * & 0 & 0\\
\hline
0 & 0 & \;*\; & 0\\
0 & 0 & 0 & *
\end{array}\right]
_{ab}.
\end{equation}
Hence the Ricci tensor will algebraically commute with the Killing tensor \emph{via} matrix multiplication: $R^{a}{}_{b}K^{b}{}_{c}=K^{a}{}_{b}R^{b}{}_{c}$.

These observations further reinforce the fact that this variant of the Lense--Thirring spacetime does indeed simplify to Schwarzschild spacetime in the appropriate  limit. 
A quick ansatz for understanding the genesis of our variant of the Lense--Thirring spacetime is to simply take Schwarzschild spacetime and subject both the the line\break element and Killing tensor to the replacement (\emph{not} a coordinate transformat\-ion):
\begin{equation}
\d\phi \longrightarrow  \left(\d\phi - {2J\over r^3} \d t\right).
\end{equation}
We shall soon use this Killing tensor to construct a Carter constant for our variant of the Lense--Thirring spacetime, but will first briefly digress to discuss Killing--Yano tensors. 

\section{Two-form square root of the Killing Tensor}

Interestingly, it is not too difficult to find a 2-form `square root' of this Killing tensor, in the sense of finding an antisymmetric tensor satisfying  $K_{ab} = - f_{ac} \, g^{cd} \, f_{db}$.
Explicitly one finds
\begin{equation}
f_{ab}= \sin\theta 
\begin{bmatrix}
0 & 0 & 2J & 0\\
0 & 0 & 0 & 0\\
-2J & 0 & 0&  r^3\\
0 & 0 & -r^3 & 0
\end{bmatrix}
_{ab} \ .
\end{equation}
We can also write this as
\begin{equation}
f_{ab} \; \d x^a \wedge \d x ^b = 
r^3 \sin\theta \;\;  \left\{\d\theta \wedge  \left(\d\phi -{2J\over r^3} \d t\right)\right\}\ .
\end{equation}
The contravariant components are even simpler
\begin{equation}
f^{ab}= {1\over r\sin\theta}
\begin{bmatrix}
0 & 0 & 0 & 0\\
0 & 0 & 0 & 0\\
0 & 0 & 0&  1\\
0 & 0 & -1 & 0
\end{bmatrix}_{ab} \ .
\end{equation}
In the orthonormal basis one finds
\begin{equation}
f_{\hat a\hat b}= r 
\begin{bmatrix}
0 & 0 & 0 & 0\\
0 & 0 & 0 & 0\\
0 & 0 & 0&  1\\
0 & 0 & -1 & 0
\end{bmatrix}_{\hat a\hat b} \ .
\end{equation}
Unfortunately while the $2$-form $f_{ab}$ is indeed a square root of the Killing tensor $K_{ab}$, it \emph{fails} to be a Killing--Yano tensor; it is at best a ``would be'' Killing--Yano tensor. Specifically, although the vector $g^{bc} f_{ab;c}=0$, which in form notation can be written as $\delta f = 0$, the $3$-index tensor  $f_{a(b;c)}$ is nonzero:
\begin{equation}
f_{a(b;c)} \; \d x^a \d x^b \d x^c = {3J\sin\theta\over 2r} 
\left\{ \d r\otimes(\d t\otimes \d\theta+\d\theta\otimes \d t ) 
- \d t\otimes(\d r\otimes \d\theta+\d\theta\otimes \d r ) \right\}.
\end{equation}
Unfortunately there does not seem to be any way to further simplify this result. 

It is also potentially worthwhile to note
\begin{equation}
f_{[ab;c]} = \epsilon_{tabc}; \qquad \hbox{equivalently} \qquad 
\d f = 3 \; \d r \wedge\d\theta\wedge\d\phi.
\end{equation}
Indeed, one sees $\delta \,\d f \propto * \d * \d f = 3 * \d * (\d r \wedge\d\theta\wedge\d\phi) 
= 3* \d (\d t) = 0$. 

Consequently, since both $\delta \d f =0$ and $\delta f=0$, we see that the $2$-form $f$ is harmonic: $\Delta f =
(\delta \d + \d\delta) f =0$. While the 2-form  $f$ is  not a Killing--Yano tensor it certainly satisfies other interesting properties.

The non-existence of the Killing--Yano tensor in turn implies the non-existence of the full Killing tower. 
When possible to do so,  one defines a \emph{principal} tensor $h_{ab}$ as the foundation of the Killing tower by demanding the existence of a 2-form $h$ such that~\cite[see discussion near page 47]{Frolov:2017}:
\begin{equation}
    \nabla_{a}h_{bc} = \frac{1}{3}\left[g_{ab}\nabla^{d}h_{dc}-g_{ac}\nabla^{d}h_{db}\right] \ .
\end{equation}
The existence of such an object is dependent upon the satisfaction of a specific integrability condition, which directly implies the spacetime be of Petrov type D. However, in reference~\cite{4-of-us}, the current authors found that the Painlev\'{e}--Gullstrand form of Lense--Thirring is Petrov type I, \emph{i.e.} not algebraically special. It follows that no principal tensor can exist for this candidate spacetime, and hence there is no associated Killing--Yano tensor. Similar oddities have also cropped up in other contexts. In references~\cite{bb-Kerr,bb-KN} those authors found that rotating black bounce spacetimes possess a nontrivial Killing tensor, and a 2-form square root thereof, but that this 2-form square root failed to be a Killing-Yano tensor.

One can also infer the non-existence of the Killing tower as a side effect of the fact that the Painlev\'{e}--Gullstrand form of Lense--Thirring does \emph{not} mathematically fall into Carter's ``off shell" $2$-free-function distortion of the Kerr spacetime~\cite[see discussion near page 42]{Frolov:2017}. 

\section{Separability of the Klein--Gordon equation}

Generally, the existence of a nontrivial Killing tensor is by itself not quite enough to guarantee separability of the Klein--Gordon equation. An explicit check needs to be carried out. There are two ways of proceeding --- either \emph{via} direct calculation, or indirectly by studying the commutativity properties of certain differential operators. We find it most illustrating to first perform a direct calculation, and then subsequently put the discussion into a more abstract framework.

\enlargethispage{20pt}
We are interested in the behaviour of the massive or massless minimally coupled Klein--Gordon equation (wave equation with possibly a mass term):
\begin{equation}
{1\over\sqrt{-g}} \partial_a \left( \sqrt{-g}\, g^{ab} \, \partial_b \Phi(t,r,\theta,\phi) \right) = \mu^2 \Phi(t,r,\theta,\phi).
\end{equation}
First we note that $\sqrt{-g} = r^2 \sin\theta$. Second, in view of the explicit Killing symmetries in the $t$ and $\phi$ coordinates we can immediately write $\Phi(t,r,\theta,\phi)  \longrightarrow \Phi(r,\theta) e^{-i\omega t} e^{in\phi}$. 

Then we are reduced to considering
\begin{equation}
\label{E:KG2}
\partial_a \left( r^2\sin\theta\, g^{ab} \, \partial_b [\Phi(r,\theta)  e^{-i\omega t} e^{in\phi}]\right) 
= \mu^2 \, r^2 \sin\theta \, \Phi(r,\theta)   e^{-i\omega t} e^{in\phi}.
\end{equation}
Now going from the Painlev\'e--Gullstrand form of the metric to Boyer--Lindquist form involves a coordinate change: $t \longleftrightarrow t + f(r)$. Under such a coordinate change $e^{-i\omega t} \longleftrightarrow  e^{-i\omega [t+f(r)]} = e^{-i\omega f(r)} e^{-i\omega t}$. 
Thence separability of the wave equation is unaffected by this coordinate transformation. Note also that the metric determinant, $\sqrt{-g} = r^2\sin\theta$, is the same in both coordinate systems.

Consequently, without loss of generality we may work in Boyer--Lindquist form, and for our current purposes it is advantageous to do so. The inverse metric is given by equation (\ref{E:BL-form})
\begin{equation}
(g^{ab})_{BL} = \left[ \begin{array}{c|cc|c}
-{1\over 1-2m/ r} &  0 & 0 & -{2J\over r^3(1-2m/r)}\\
\hline
0 & 1-{2m\over r} & 0 & 0\\
0& 0 & {1\over r^2} & 0\\
\hline
-{2J\over r^3(1-2m/r)} & 0 & 0 & {1\over r^2\sin^2\theta} - {4J^2\over r^6(1-2m/r)}\\
\end{array}
\right]^{ab}.
\end{equation}
Then the Klein--Gordon equation (\ref{E:KG2}) reduces to
\begin{eqnarray}
&&\sin\theta \; \partial_r [ r^2 (1-2m/r) \partial_r\Phi] + \partial_\theta [\sin\theta\; \partial_\theta\Phi] 
\nonumber\\
&&\qquad 
+r^2\sin\theta\left( {\omega^2 \over 1-2m/r} - {4Jn\omega\over r^3(1-2m/r)} 
- n^2\left[{1\over r^2\sin^2\theta} - {4J^2\over r^6(1-2m/r)}\right] \right) \Phi
\nonumber\\
&&\qquad = \mu^2 r^2 \sin\theta \; \Phi.
\end{eqnarray}
That is
\begin{eqnarray}
&&{\partial_r [ r^2 (1-2m/r) \partial_r\Phi]} 
 + {\partial_\theta [\sin\theta\; \partial_\theta\Phi] \over \sin\theta}
 - {n^2\over\sin^2\theta} \Phi
+r^2\left( {(\omega - 2Jn/r^3)^2 \over 1-2m/r} \right) \Phi
\nonumber\\
&&\qquad = \mu^2 r^2 \Phi.
\end{eqnarray}
This is now manifestly separable:
\begin{eqnarray}
&&{\partial_r [ r^2 (1-2m/r) \; \partial_r\Phi]} 
+r^2 \; {(\omega - 2Jn/r^3)^2 \over 1-2m/r} \Phi
- \mu^2 r^2 \Phi
\nonumber\\
&&\qquad = 
-{\partial_\theta [\sin\theta\; \partial_\theta\Phi] \over \sin\theta}
 + {n^2\over\sin^2\theta} \Phi. 
\end{eqnarray}
\def\R{{\mathcal{R}}}
To be even more explicit about this, let us write $\Phi(r,\theta) = \R(r) \Theta(\theta)$, then:
\begin{eqnarray}
&&{1\over \R(r)}\left\{ {\partial_r [ r^2 (1-2m/r) \; \partial_r \R(r)]} 
+r^2 \; {(\omega - 2Jn/r^3)^2 \over 1-2m/r} \R(r)
- \mu^2 r^2 \R(r) \right\}
\nonumber\\
&&\qquad = 
{1\over\Theta(\theta)} \left\{ -{\partial_\theta [\sin\theta\; \partial_\theta\Theta(\theta)] \over \sin\theta}
 + {n^2\over\sin^2\theta} \Theta(\theta) \right\}. 
\end{eqnarray}
(The left-hand-side depends only on $r, \R(r)$, and its derivatives; the right-hand-side depends only on $\theta, \Theta(\theta)$ and its derivatives.)
So we have explicitly verified that the massive Klein--Gordon equation (the wave equation) does in fact separate on our variant of the Lense--Thirring spacetime.

A more abstract way of checking for separability of the wave equation is to consider the commutativity properties of appropriate  differential operators. 
Assume one has a nontrivial Killing tensor $K_{ab}$, and define the Carter differential operator $\K$ and wave differential operator $\Box$ by:
\begin{equation}
\K \Phi = \nabla_a (K^{ab} \nabla_b \Phi); \qquad \Box \Phi = \nabla_a (g^{ab} \nabla_b \Phi) \ .
\end{equation}
Then a brief (but somewhat messy) calculation yields:
\begin{equation}
[\K,\Box]\Phi = {2\over3} 
\left( \nabla_d [R,K]^d{}_b   \right)  \nabla^b \Phi \ .
\end{equation} 
(See proposition 1.3 of the recent reference~\cite{Giorgi}, as modified in appendix \ref{A:A} below. See also the considerably older discussion presented in reference~\cite{Benenti}.)

Then a \emph{necessary and sufficient} condition for the Carter operator to commute with the wave operator is that 
\begin{equation}
 \nabla_d [R,K]^d{}_b  =0 \ .
\end{equation}
Since, as we have already noted, $[R,K]$ can be viewed as 2-form, the condition $ \nabla_d [R,K]^d{}_b  =0 $ can be recast in the notation of differential forms as $\delta [R,K] =0$. 
This condition is certainly satisfied for Ricci-flat and Einstein manifolds, (such as Kerr and Kerr--de~Sitter), but a weaker (yet still sufficient) condition is the vanishing of the commutator $[R,K]^d{}_b  =0$, and we have already seen that this commutator vanishes for our variant of Lense--Thirring spacetime.\footnote{This tensor commutator also vanishes for Kerr--Newman spacetimes, and for the black-bounce modifications of Kerr and Kerr--Newman spacetimes studied in \cite{bb-Kerr, bb-KN}.
Thus the wave equation is separable on all of these spacetimes.} This is enough to imply separability of the wave equation on our variant of Lense--Thirring spacetime.

\section{Carter constant and other conserved quantities}

Extraction of the (generalized) Carter constant is now straightforward:
\begin{equation} 
\mathcal{C}=K_{ab}\frac{\d x^a}{\d\lambda}\frac{\d x^b}{\d\lambda}=r^4\left[ \left(\frac{\d\theta}{\d\lambda}\right)^2 + \sin^2\theta \left(\frac{\d\phi}{\d\lambda}-\frac{2J}{r^3}\frac{\d t}{\d\lambda}\right)^2 \right] \ ,
\end{equation}
for any affine parameter $\lambda$. Without loss of generality we may enforce that $\lambda$ be future-directed, as is conventional. Note that by construction we have $\C \geq 0$.

In addition to the Carter constant, we have three other conserved quantities:
\begin{equation}\label{E}
E = -\xi_a\dfrac{\d x^a}{\d\lambda} = \left(1-\frac{2m}{r}-\frac{4J^2\sin^2\theta}{r^4} \right)\frac{\d t}{\d\lambda}-\sqrt{\frac{2m}{r}}\frac{\d r}{\d\lambda}+\frac{2J\sin^2\theta}{r}\frac{\d\phi}{\d\lambda} \ ;
\end{equation}
\begin{equation} \label{L}
L = \psi_a\frac{\d x^a}{\d\lambda} = r^2\sin^2\theta\frac{\d\phi}{\d\lambda} -\frac{2J\sin^2\theta}{r}\frac{\d t}{\d\lambda} \ ,
\end{equation}
and
\begin{equation}\label{Geo_eqn}
\begin{split}
\epsilon=g_{ab}\frac{\d x^a}{\d\lambda}\frac{\d x^b}{\d\lambda}= & -\left(\frac{\d t}{\d\lambda}\right)^2+\left(\frac{\d r}{\d\lambda}+\sqrt{\frac{2m}{r}}\frac{\d t}{\d\lambda}\right)^2\\
& + r^2\left[\left(\frac{\d\theta}{\d\lambda}\right)^2+\sin^2\theta\left(\frac{\d\phi}{\d\lambda}-\frac{2J}{r^3}\frac{\d t}{\d\lambda}\right)^2\right] \ .
\end{split}
\end{equation}
The conserved quantities $E$ and $L$ arise from the time translation and azimuthal Killing vectors, respectively given by $\xi^a=(1;0,0,0)^a$ and  $\psi^a=(0,0,0,1)^a$.  In contrast the conserved quantity $\epsilon$, with $\epsilon \in \{0,-1\}$ for null and timelike geodesics respectively, arises from the \emph{trivial} Killing tensor $g_{ab}$. 
\enlargethispage{20pt}

Note that if $\epsilon=0$ then, without  loss of generality, we can rescale the affine parameter $\lambda$ to set \emph{one} of the constants $\{\C,E,L\} \to 1$. It is intuitive to set $E\to 1$.
In contrast if $\epsilon=-1$ then $\lambda = \tau$ is the proper time and there is no further freedom to rescale the affine parameter. $E$ then has real physical meaning and the qualitative behaviour is governed by the \emph{sign} of $E^2+\epsilon$. Concretely: Is $E <1$ (bound orbits), is $E=1$ (marginal orbits), or is $E>1$ (unbound orbits)?

We can now greatly simplify these four conserved quantities by rewriting them as follows:
\begin{equation} \label{L_2}
L=r^2\sin^2\theta \left(\frac{\d\phi}{\d\lambda}-\frac{2J}{r^3}\frac{\d t}{\d\lambda}\right) \ ;
\end{equation}
\begin{equation} \label{C_2}
\mathcal{C}=r^4\left(\frac{\d\theta}{\d\lambda}\right)^2 +{L^2\over \sin^2\theta} \ ;
\end{equation}
\begin{equation}\label{Geo_eqn_2}
\epsilon=  -\left(\frac{\d t}{\d\lambda}\right)^2+\left(\frac{\d r}{\d\lambda}+\sqrt{\frac{2m}{r}}\frac{\d t}{\d\lambda}\right)^2 +{\mathcal{C}\over r^2} \ ;
\end{equation}
\begin{equation}\label{E_2}
E=\left(1-\frac{2m}{r}\right)\frac{\d t}{\d\lambda} - \sqrt{\frac{2m}{r}}\frac{\d r}{\d\lambda} + \frac{2J}{r^3}L \ .
\end{equation}
Notice that by construction $\C \geq L^2$. Furthermore, the form of the Carter constant, equation \eqref{C_2}, gives a range of forbidden declination angles for any given, non-zero values of $\mathcal{C}$ and $L$. 

We require that $\d\theta/\d\lambda$ be real, and from equation \eqref{C_2} this implies the following requirement:
\begin{equation}
\left(r^2\frac{\d\theta}{\d\lambda}\right)^2=\mathcal{C}-\frac{L^2}{\sin^2\theta}\geq 0
\quad\Longrightarrow\quad
\sin^2 \theta \geq {L^2\over \C} \ .
\end{equation}
Then provided $\C \geq L^2$, which is automatic in view of \eqref{C_2}, we can define $\theta_* \in [0,\pi/2]$ by setting 
\begin{equation}
\theta_* = \sin^{-1}(|L|/\sqrt{\C}) \ .
\end{equation}
 Then the allowed range for $\theta$ is the equatorial band
\begin{equation}
\label{E:theta_range}
\theta \in \Big[ \theta_*, \pi -\theta_*\Big] \ .
\end{equation}
For $L^2=\C$ we have $\theta=\pi/2$; the motion is restricted to the equatorial plane.\\
For $L=0$ with $\C>0$ the range of $\theta$ is \emph{a priori} unconstrained; $\theta\in[0,\pi]$.\\
For $L=0$ with $\C=0$ the declination is fixed $\theta(\lambda)=\theta_{0}$, and the motion is restricted to a constant declination conical surface.

Using equations \eqref{L_2}, \eqref{C_2}, \eqref{Geo_eqn_2} \& \eqref{E_2} we can (at least in principle) analytically solve for the four unknown functions $\d t/\d\lambda$, $\d r/\d\lambda$, $\d\theta/\d\lambda$ and $\d\phi/\d\lambda$ as explicit functions of $r$ and $\theta$, parameterized by the four conserved quantities $\C$, $E$, $L$, and $\epsilon$, as well as the quantities $m$ and $J$ characterizing mass and angular momentum of the central object. The resulting formulae are quite tedious and will be reported elsewhere. 

\section{Conclusions}

\enlargethispage{20pt}
From the discussion above we have seen that it is relatively straightforward to find a non-trivial Killing tensor for the Painlev\'e--Gullstrand version of the Lense--Thirring spacetime.
We have also demonstrated separability of the Klein--Gordon equation, and the \emph{non-existence} of a Killing--Yano 2-form.
Once we have found the non-trivial Killing tensor, we can easily extract the Carter constant; the fourth constant of the motion. Then the geodesic equations become integrable, which allows us (in principle) to solve for myriads of general geodesics. 

\section*{Acknowledgements}

JB was supported by a MSc scholarship funded by the Marsden Fund, 
via a grant administered by the Royal Society of New Zealand.
\\
TB was supported by a Victoria University of Wellington MSc scholarship, 
and was also indirectly supported by the Marsden Fund, 
via a grant administered by the Royal Society of New Zealand.
\\
AS was supported by a Victoria University of Wellington PhD Doctoral Scholarship,
and was also indirectly supported by the Marsden fund, 
via a grant administered by the Royal Society of New Zealand.
\\
MV was directly supported by the Marsden Fund, \emph{via} a grant administered by the Royal Society of New Zealand.

\appendix
\section{Wave operators}
\label{A:A}

\enlargethispage{40pt}
In the recent reference \cite[page 9, proposition 1.3]{Giorgi}  the author demonstrated that 
\begin{eqnarray}
[\K,\Box]\Phi &=& \left\{ \left(\nabla_c R - {4\over3} \nabla_d R^d{}_c \right)   K^c{}_b \right.
\nonumber\\
&&
\left.
+ {2\over3}\left( R^{dc} \nabla_d K_{cb} - R^c{}_b\nabla_d K^d{}_c 
- \{\nabla_c R^ d{}_b \} K^c{}_d \right) \right\} \nabla^b \Phi \ .
\end{eqnarray}
Now use the (twice contracted) Bianchi identity, in the opposite direction from what one might expect, to temporarily make things more complicated:
\begin{equation}
\nabla_c R = 2\nabla_d R^d{}_c \ .
\end{equation}
Then  proposition 1.3 becomes
\begin{equation}
[\K,\Box]\Phi = \left\{ \left( + {2\over3} \nabla_d R^d{}_c \right)   K^c{}_b 
+ {2\over3}\left( R^d{}_c \nabla_d K^c{}_b - R^c{}_b\nabla_d K^d{}_c 
- \{\nabla_c R^ d{}_b \} K^c{}_d \right) \right\} \nabla^b \Phi \ .
\end{equation}
That is
\begin{equation}
[\K,\Box]\Phi = {2\over3} 
\left\{ 
R^d{}_c \nabla_d K^c{}_{b} - R^c{}_b\nabla_d K^d{}_c 
- (\nabla_c R^ d{}_b ) K^c{}_d  +  (\nabla_d R^d{}_c)    K^c{}_b 
\right\} \nabla^b \Phi \ .
\end{equation}
Relabelling some indices
\begin{equation}
[\K,\Box]\Phi = {2\over3} 
\left\{ 
R^d{}_c \nabla_d K^c{}_{b} - R^c{}_b\nabla_d K^d{}_c 
- \{\nabla_d R^c{}_b \} K^d{}_c  +  \{\nabla_d R^d{}_c\}    K^c{}_b 
\right\} \nabla^b \Phi \ .
\end{equation}
That is
\begin{equation}
[\K,\Box]\Phi = {2\over3} 
\nabla_d \left\{ R^d{}_c  K^c{}_{b} - K^d{}_c  R^c{}_{b} \right\}  \nabla^b \Phi \ .
\end{equation}
Finally rewrite this as:
\begin{equation}
[\K,\Box]\Phi = {2\over3} 
\left( \nabla_d [R,K]^d{}_b   \right)  \nabla^b \Phi \ .
\end{equation}
(See also the considerably older discussion in reference~\cite{Benenti}, using somewhat different terminology.)

\bigskip
\hrule
\hrule


\end{document}